# Visual tool for estimating the fractal dimension of images

I. V. Grossu [1], C. Besliu [1], M.V.Rusu[1], Al. Jipa [1], C. C. Bordeianu [1], D. Felea [2], E.Stan[2], T. Esanu[1]

[1] *University of Bucharest, Faculty of Physics, Bucharest-Magurele, Romania*
[2] *Institute of Space Sciences, Bucharest-Magurele, Romania*

**Abstract**

This work presents a new Visual Basic 6.0 application for estimating the fractal dimension of images, based on an optimized version of the box-counting algorithm. Following the attempt to separate the real information from "noise", we considered also the family of all band-pass filters with the same band-width (specified as parameter). The fractal dimension can be thus represented as a function of the pixel color code. The program was used for the study of paintings cracks, as an additional tool which can help the critic to decide if an artistic work is original or not. In its second version, the application was extended for working also with comma separated values files and three-dimensional images.

## Program summary

*Manuscript title: Visual tool for estimating the fractal dimension of images*
*Authors: I.V.Grossu, C. Besliu, Al. Jipa, C.C. Bordeianu, D. Felea*
*Program title: Fractal Analysis v01*
*Licensing provisions:*
*Programming language: MS Visual Basic 6.0*
*Computer(s) for which the program has been designed: PC*
*Operating system(s) for which the program has been designed: MS Windows 98 or later*
*RAM required to execute with typical data: 30M*
*Has the code been vectorised or parallelized: no*
*Number of processors used: 1*
*Supplementary material:*
*Keywords: fractal, box-counting*
*PACS:*
*CPC Library Classification:*
*External routines/libraries used:*
*Nature of problem: Estimating the fractal dimension of images.*
*Solution method: Optimized implementation of the box-counting algorithm. Use of a band-pass filter for separating the real information from "noise". User friendly graphical interface.*
*Restrictions: Although various file-types can be used, the application was mainly conceived for the 8-bit grayscale, windows bitmap file format.*
*Additional comments:*
*Running time: In a first approximation, the algorithm is linear.*

## 1. Introduction

Based on the box-counting algorithm and the band-pass filter, we developed a set of tools for calculating the fractal dimension of images. Microsoft Visual Basic 6.0 [1] was chosen because it provides an easy to use environment for fast development of windows forms applications. The "Pckage and deployment wizard" simplifies also the deployment, by automatically generating setup packages.

In order to improve the application performances, an important attention was payed to the algorithms optimization. Our efforts were focused also on creating an easy to exploit, user friendly graphical interface.

## 2. Application description

Although various file-types are supported, the program was mainly conceived for the 8-bit grayscale, windows bitmap file format. In order to minimize the processing time, the entire image is loaded into one array of integers.

Our first goal was to extract the real information from the parasite background, generated by various unwanted effects. One very simple method we implemented in the application is the band-pass filter. The lower and the upper values for this functionality can be specified either directly, if the color codes are known, either by clicking on the desired region inside the image.

For computing the fractal dimension, we employed the Box-Counting algorithm [2]. In this context, the generalized dimension $d_f$ can be deduced as:

$$d_f = \lim_{r \to 0} \frac{\log N(r)}{\log(1/r)} \quad (1)$$

where $N(r)$ represents the number of boxes, with length $r$, needed to cover the object. In our program, this method can be used only in conjunction with the band-pass filter, which "converts" the bitmap into a boolean array. The proposed solution is based on counting all boxes containing at least one point in the selected band.

The minimum box length is limited, by the finite resolution, to one pixel. For verifying the existence of limit (1), the length $r_n$ is doubled on each iteration, and the number of boxes containing information $N(r_n)$ is stored in one array. For simplicity, we restricted the analysis only to the image regions wich sizes can be expressed as powers of two. Thus, the fractal dimension can be deduced from the slope of the line obtained by applying the least squares method for the set of all points $(log (1/ r_n), log N(r_n))$.

One can notice that, following the previous described technique, it is possible to calculate also, on each iteration $n$, the number of boxes $N(r_{n+1})=N(2r_n)$, corresponding to the next step:

```
       For i = 0 To WMax Step 2
          For j = 0 To HMax Step 2
             'bNext is a boolean used to get the solution for the next level
             If mLevel(i, j) Then
                lngNoBox = lngNoBox + 1
                bNext  = True
             End If
             If mLevel(i + 1, j) Then
                lngNoBox = lngNoBox + 1
                bNext  = True
             End If
             If mLevel(i, j + 1) Then
                lngNoBox = lngNoBox + 1
                bNext  = True
             End If
             If mLevel(i + 1, j + 1) Then
                lngNoBox = lngNoBox + 1
                bNext  = True
             End If
             'mLevel is an array of booleans used to store the solution for each level
             mLevel(Int(i / 2), Int(j / 2)) = bNext
          Next j
       Next i
```

The algorithm complexity is thus reduced to:

$$C = w \times h \left( 1 + \frac{1}{4} + \frac{1}{4^2} + \cdots + \frac{1}{4^{\log_2(Min(w,h))}} \right) \quad (2)$$

where *w* and *h* represent the width, respectively the height of the selected image region.

Another type of analysis is based on considering the family of all band-pass filters with the same band-with. The fractal dimension can be represented, in real time, as a function of the pixel color code (the lower value corresponding to each band-pass member). In this context, the existence of a plate could be an important criterion which can help on better separating the real information from noise.

In its second version, the application was extended for working with data stored in comma separated values (csv) files. In this context, we implemented also a three dimensional version of the, previous discussed, box-counting algorithm.

## 3. Conclusion

We developed a set of, easy to use, visual tools for estimating the fractal dimension of images. In order to obtain a good response time, an important attention was pay to the algorithms optimizations. The installation is effortless (automatically generated setup package), and the application requires minimal resources to run (CPU 133MHz, 256M RAM, MS Windows 98 or later).

The program was used, for example, in a student laboratory for calculating the fractal dimensions corresponding to a set of Helle-Shaw cells with lifting panes [3,4] and acrylic colors (Fig.1).

In the last years, the world of art is facing a lot of new falsification techniques. Thus, for reproducing the fractures, which naturally appear in old paintings, the copies are sometimes scratched or exposed to high temperatures [5]. One idea could be to use the presented software for the fractal analysis of cracks [3,6], as an additional tool which can help the critic deciding if an artistic work is original or not. Further analysis along those lines is currently in progress.

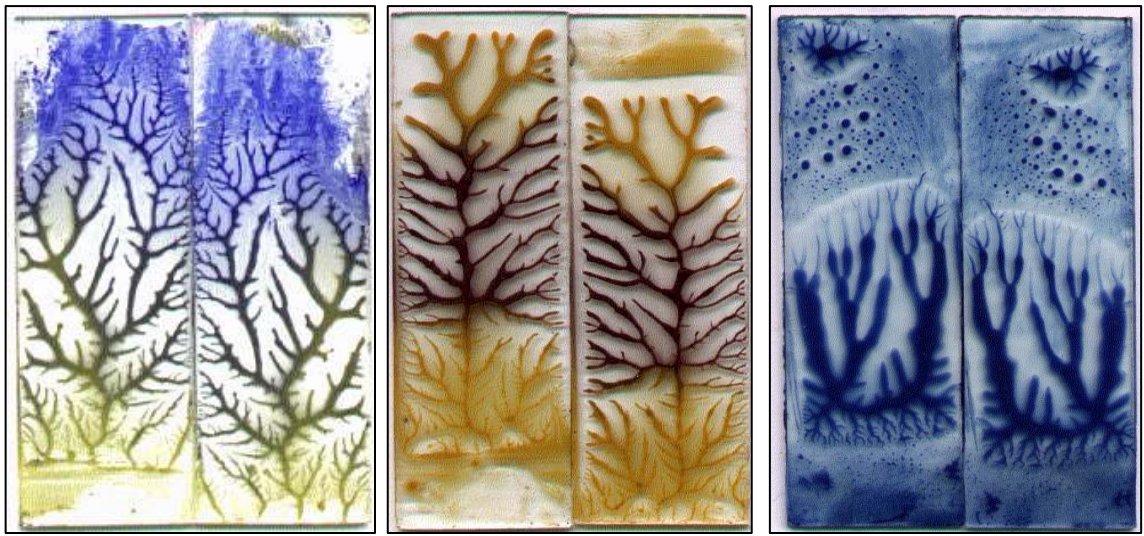

Fig. 1. Helle-Shaw cells with lifting panes.

**References**

[1]  F. Balena, Programming Microsoft Visual Basic 6.0, Microsoft Press, U.S., 1999
[2]  R.H. Landau, M.J. Paez and C.C. Bordeianu, Computational physics : Problem solving with computers, Wiley-VCH-Verlag, Weinheim, 2007, pp. 293-306
[3]  I.V.Grossu, M.V.Rusu, A.Teodosiu, Fractals in a particular process. Fractals in the investigation of artistic works., National Conference of Physics, Romania, Constanta 21-23 September 2000
[4]  Kabiraj and Tarafdar, Physica A328, 305(2003)
[5]  A. Teodosiu, Din universul ascuns al operei de arta, Allfa, Romania, 2001, pp. 113-122
[6]  C. Allain, F. Parisse, La Recherche 288 Juin 1996, Des fractures bien rangees, pp. 50